\documentclass[journal,10pt]{IEEEtran}
\usepackage{amsmath,graphicx,cite,color}
\usepackage[ruled,linesnumbered]{algorithm2e}
\usepackage{amsfonts,amssymb}
\usepackage{multirow}
\usepackage{balance}

\IEEEoverridecommandlockouts

\newcommand{\eqdef}{\stackrel{\triangle}{=}}
\newcommand{\A}{{\bf A}}

\newcommand{\z}{{\bf z}}
\newcommand{\x}{{\bf x}}

\newcommand{\g}{{\bf g}}

\renewcommand{\a}{{\bf a}}
\newcommand{\w}{{\bf w}}

\newcommand{\HH}{{\bf H}}
\newcommand{\wCx}{{\bf C}_{\bf x}}
\newcommand{\mi}{\mathrm{i}}
\newcommand{\bnabla}{\boldsymbol{\nabla}}

\begin{document}


\title{Blind Capon Beamformer Based on Independent Component Extraction: Single-Parameter Algorithm}

\author{\authorblockN{Zbyn\v{e}k Koldovsk\'{y}\authorrefmark{1}, Jaroslav \v{C}mejla\authorrefmark{1}, and Stephen O'Regan\authorrefmark{2}
}\\
\authorblockA{\authorrefmark{1}
Acoustic Signal Analysis and Processing Group, 
Technical University of Liberec, Czech Republic.}\\
\authorblockA{\authorrefmark{2}Naval Surface Warfare Center Carderock Division, West Bethesda, Maryland, USA}
\thanks{This work was supported by US Office of Naval Research Global through Project No.~N62909-23-1-2084.} 
}

\maketitle
\begin{abstract}
We consider a phase-shift mixing model for linear sensor arrays in the context of blind source extraction. We derive a blind Capon beamformer that seeks the direction where the output is independent of the other signals in the mixture. 
The algorithm is based on Independent Component Extraction and imposes an orthogonal constraint, thanks to which it optimizes only one real-valued parameter related to the angle of arrival.
The Cram\'er-Rao lower bound for the mean interference-to-signal ratio is derived. The algorithm and the bound are compared with conventional blind and direction-of-arrival estimation+beamforming methods, showing improvements in terms of extraction accuracy. An application is demonstrated in frequency-domain speaker extraction in a low-reverberation room.
\end{abstract}
\begin{keywords}
	Blind Source Separation, Blind Source Extraction, Independent Component Analysis, Independent Vector Analysis, Speaker Extraction
\end{keywords}
%

\section{Introduction}
\label{sec:intro}
Multichannel source extraction is a fundamental task in array signal processing where the goal is to retrieve a particular signal of interest (SOI) from mixtures of signals observed by sensors. The problem is challenging because of the lack of information about the SOI, its direction of arrival (DOA), the periods of activity, the noise covariance matrix, and similar important knowledge that is obscured in the observed signal mixture. Spatial filters provide efficient tools but require some of the above information to be available \cite{vanveen1988,elbir2023}. In Blind Source Extraction (BSE), the aim is to solve the problem without such knowledge, based on general principles \cite{comon1994}. This letter addresses BSE through Independent Component Extraction (ICE), which is based on the assumption that the SOI is statistically independent of the other signals in the mixture
\cite{comon2010handbook,koldovsky2019TSP}.



The source extraction problem is described by the (complex-valued) linear instantaneous mixing model
\begin{equation}\label{modelICE}
\x(n) = \a s(n) + {\bf y}(n),
\end{equation} 
where $n$ is the sample index $n=1,\dots,N$, $\x(n)$ is the $d\times 1$ vector of observed signals, and $s(n)$ and ${\bf y}(n)$ denote, respectively, the SOI and the other signals; $\a$ is the $d\times 1$ mixing vector whose elements correspond to weights with which the SOI is received by the sensors; $\a$ can have arbitrary scale due to the scaling ambiguity (see, e.g., Section II.B in \cite{koldovsky2019TSP}.) In the beamforming literature, much attention has been paid to structured $\a$ reflecting spatial properties of the SOI and various array geometries \cite{vantrees2002}. In this paper, we consider the far-field mixing model with
\begin{equation}\label{structureda}
    \a(\lambda)=\begin{bmatrix}
        1 & e^{\mi\lambda v_2} & \dots & e^{\mi\lambda v_{d}}
    \end{bmatrix}^T,
\end{equation}
where $\lambda$ is a real-valued parameter, and $v_j$, $j=2,\dots,d$ are known weights collected in the vector ${\bf v}=[0, v_2, \dots,v_{d}]^T$; $\mi$ denotes the imaginary unit; $\cdot^T$ denotes the transposition. The model includes the important case where the SOI reaches a linear array of sensors as a plane wave without reflections so that its images on sensors differ only by phase shifts. The model has potential applications where the phase shifts depend almost exclusively on the DOA, such as in speaker extraction in low-reverberant acoustic environments or in radio-electronic source detection and tracking.

While, in the beamforming theory, this has been a well-studied problem for decades (see, e.g. \cite{vantrees2002,benesty2016,elbir2023}), the model has not yet been considered in the context of ICE. The conventional model \eqref{modelICE} describes multi-source and multi-path environments and ad hoc (unknown) sensor arrangements, which leads to general-purpose methods. Structured models such as \eqref{structureda} enable us to employ knowledge {  such as} the array geometry, which can enhance the potential of ICE for DOA estimation and beamforming operating at lower signal-to-interference ratio levels than conventional methods. 


In this letter, we derive an ICE algorithm based on the structured model. An orthogonal constraint is imposed whose analytic form is the same as that of the minimum power distortionless beamformer (MPDR) \cite{vantrees2002}. The algorithm optimizes only one real-valued parameter and can be seen as a blind Capon beamformer \cite{elbir2023} performing joint DOA estimation and source extraction.
We also derive the Cram\'er-Rao-induced lower bound for the mean interference-to-signal ratio to reveal the theoretical best extraction accuracy achievable through the model. In simulations, the algorithm and the bound are compared with conventional DOA+beamforming approaches and general-purpose ICE. A straightforward broadband extension of the algorithm, based on Independent Vector Extraction (IVE), is demonstrated for frequency-domain speaker extraction in a low-reverberation room.

\section{Algorithm Development}
Let $\w$ be a separating (beamforming) vector such that $s(n) = \w^H\x(n)$; $\cdot^H$ is the conjugate transposition. 
ICE is solved by jointly estimating $\a$ and $\w$ from $\x(n)$ based on the statistical model \cite{koldovsky2019TSP}; the vectors are linked through the distortionless condition $\w^H\a=1$. The existence of $\w$ such that $s(n) = \w^H\x(n)$ is ensured through the so-called determined model, in which the observation model \eqref{modelICE} can be written in the form $\x(n)=\A[s(n)\,\,\z(n)^T]^T$ where $\A$ is a square mixing matrix whose first column is $\a$. Hence, ${\bf y}(n)$ is assumed to belong to a $(d-1)$-dimensional subspace of so-called background signals denoted by the $(d-1)\times 1$ vector $\z(n)$ \cite{kitamura2016determined,koldovsky2019TSP}. 

\subsection{Statistical model and contrast function}
We will model each signal as a sequence of identically and independently distributed zero-mean random variables (therefore, the argument $n$ can be omitted). The SOI is assumed to have a non-Gaussian pdf $p(s)$, which is not known and is later replaced by a model pdf given by 
$s \sim f\left(\frac{s}{\sigma}\right)\sigma^{-2}$,
where $\sigma^{2}$ is the variance of $s$, and $f(\cdot)$ is a suitable pdf of a normalized non-Gaussian random variable {  \cite{koldovsky2021fastdiva}}. Typically, $f(\cdot)$ need not be defined explicitly. Instead, a nonlinear function $\phi(\cdot)$ that corresponds to its score function $-\frac{\partial \log f(\cdot)}{\partial s}$ is selected, and $\nu^{-1}\phi(\cdot)$ is used in place of the score function, where 
\begin{equation}\label{nu}
    \nu= {\rm E}\left[\phi\left(\frac{{s}}{\sigma}\right)\frac{{s}}{\sigma}\right].
\end{equation} 
The normalization of $\phi(\cdot)$ by $\nu$ is needed to satisfy a general property of score functions: For any score function $\psi(s)$ it holds that ${\rm E}[s\psi(s)]=1$; {  see, e.g., Section III.A in \cite{koldovsky2019TSP}}. 

The background signals $\z$ are assumed to be zero-mean circular Gaussian having an unknown covariance matrix denoted by ${\bf C}_{\bf z}$. In principle, this assumption means that we neglect the properties of signals other than the SOI. This often leads to statistical suboptimality, which is worthwhile in terms of simplification; see discussions, e.g., in \cite{koldovsky2021fastdiva}.

The contrast function for the estimation of the parameter vectors $\a$ and $\w$ derived based on the maximum likelihood principle is given by (see Equation (13) in \cite{koldovsky2021fastdiva})
\begin{multline}\label{contastICE}
    \mathcal{C}_{{\bf w},{\bf a}}\left({\bf w},{\bf a}\right) ={\rm E}\left[\log f\left(\frac{{s}}{\sigma}\right)\right]  -\log\sigma^2 -\hat{\rm E}\left[{\bf z}^H{\bf C}_{\bf z}^{-1}{\bf z}\right]  \\ 
    + (d-2)\log |\gamma|^2, 
\end{multline}
where  $s = \w^H\x$, $\sigma^2={\rm E}[| s|^2]$, ${\bf z}={\bf B}\x$ with \mbox{${\bf B}=[\g,\,\, -\gamma{\bf I}_{d-1}]$} being the blocking matrix where $\gamma$ and $\g$ are, respectively, the upper and lower parts of $\a$, i.e., \mbox{$\a=[\gamma,\, \g^T]^T$}. ${\bf I}_d$ denotes the $d\times d$ identity matrix. The unknown ${\bf C}_{\bf z}$ is, in algorithms, replaced by the current sample-based estimate of the covariance matrix of ${\bf z}$. 

\subsection{Orthogonal constraint}
Since $s$ and ${\bf z}$ are independent, they are also uncorrelated. It is thus possible to impose the so-called orthogonal constraint (OC), which requires that the subspace generated the current estimate of ${s}$ is orthogonal to that of ${\bf z}$. Specifically, together with the distortionless condition $\w^H\a=1$, the OC is that ${\rm E}[{\bf z}s^*]={\bf B}\wCx{\bf w}^H=0$ where $\wCx={\rm E}[\x\x^H]$ is the (sample-based) covariance matrix of $\x$. As shown in \cite{koldovsky2019TSP}, with the OC, it is possible to express $\w$ as a function of $\a$ 
\begin{equation}\label{orthogonalconstraint}
	{\bf w}(\a)=\frac{\wCx^{-1}\a}{\a^H\wCx^{-1}\a}=\sigma^2\wCx^{-1}\a,
\end{equation}
where it holds that $\sigma^2=(\a^H\wCx^{-1}\a)^{-1}=\w^H\wCx\w$; similarly, $\a(\w)=\sigma^{-2}\wCx\w$ when $\w$ is treated as the free variable. We can now introduce new contrast functions $\mathcal{C}_{\bf a}(\a)=\mathcal{C}_{{\bf w},{\bf a}}(\sigma^2\wCx^{-1}\a,\a)$ and $\mathcal{C}_{\bf w}(\w)=\mathcal{C}_{{\bf w},{\bf a}}(\w,\sigma^{-2}\wCx\w)$. For the model studied in this letter, we define 
\begin{equation}\label{contrastLPS}
    \mathcal{C}_{\lambda}(\lambda)=\mathcal{C}_{\bf a}(\a(\lambda))=\mathcal{C}_{\bf w}(\w(\lambda))
\end{equation} where, owing to \eqref{orthogonalconstraint}, $\w(\lambda)=\sigma^2\wCx^{-1}\a(\lambda)$. $\mathcal{C}_{\lambda}(\lambda)$ will now be subject to maximization to find the (blind) estimate of $\lambda$.

\subsection{Second-order optimization}
{  We now develop a Newton-Raphson optimization-based algorithm similar to \cite{koldovsky2021fastdiva}. To this end, the first and second-order derivatives of $\mathcal{C}_{\lambda}$ are computed using known derivatives}
for the unstructured mixing models \cite{koldovsky2019TSP,koldovsky2021fastdiva}. By the complex-valued chain rule \cite{petersen2008} it holds that
 $   \frac{\partial\mathcal{C}_\lambda}{\partial \lambda}=
    \left(\frac{\partial\mathcal{C}_\w}{\partial \w}\right)^T
    \frac{\partial \w}{\partial \lambda}+
    \left(\frac{\partial\mathcal{C}_\w}{\partial \w^*}\right)^T
    \frac{\partial \w^*}{\partial \lambda}
    =\left(\frac{\partial\mathcal{C}_\a}{\partial \a}\right)^T
    \frac{\partial \a}{\partial \lambda}+
    \left(\frac{\partial\mathcal{C}_\a}{\partial \a^*}\right)^T
    \frac{\partial \a^*}{\partial \lambda}$. By \cite{koldovsky2019TSP}, 
we know that
\begin{align}
   \bnabla_{\bf w} &\eqdef \frac{\partial\mathcal{C}_\w}{\partial \w^*} 
   = {\bf a}(\w)-\nu^{-1}{\rm E}\left[\phi\left(\frac{{s}}{\sigma}\right)\frac{\bf x}{\sigma}\right],\\
   \bnabla_{\bf a} &\eqdef \frac{\partial\mathcal{C}_\a}{\partial \a^*} 
   =\sigma^2\wCx^{-1}\bnabla_{\bf w}\label{grada}.
\end{align}
Since \eqref{contrastLPS} is real-valued, it holds that $\bnabla_{\w^*}=\bnabla_{\w}^*$ and $\bnabla_{\a^*}=\bnabla_{\a}^*$. By \eqref{structureda},
\begin{equation}\label{diffa}
 \frac{\partial\a}{\partial \lambda} = \mi(\a\odot{\bf v})   \quad\text{and}\quad
 \frac{\partial\a^*}{\partial \lambda} = -\mi(\a^*\odot{\bf v}^*),
\end{equation}
where $\cdot^*$ denotes the conjugate value, and $\odot$ denotes the Hadamard (element-wise) product. By employing \eqref{grada} and \eqref{diffa} in the chain rule,
\begin{equation}\label{firstderivative}
    \frac{\partial\mathcal{C}}{\partial \lambda}=-2\Im\left\{\bnabla_{\bf a}^H(\a\odot{\bf v})\right\}=
    -2\hat\sigma^2\Im\left\{\bnabla_{\bf w}^H\wCx^{-1}(\a\odot{\bf v})\right\},
\end{equation}
where $\Im\{\cdot\}$ denotes the imaginary part of the argument.

Similarly, using known results, we now turn to express the second derivative of $\mathcal{C}_{\lambda}$ as
\begin{multline}\label{secondderivative}
    \frac{\partial^2\mathcal{C}}{\partial \lambda^2}=
    -2\frac{\partial\sigma^2}{\partial \lambda}\Im\left\{\bnabla_{\bf w}^H\wCx^{-1}(\a\odot{\bf v})\right\}\\
    -2\sigma^2\Im\left\{\frac{\partial \bnabla_{\bf w}^H}{\partial\lambda}\wCx^{-1}(\a\odot{\bf v})+\mi\bnabla_{\bf w}^H\wCx^{-1}(\a\odot{\bf v}^2)\right\},
\end{multline}
where ${\bf v}^2={\bf v}\odot{\bf v}$.
To express $\frac{\partial\bnabla_\w}{\partial\lambda}$, we apply the chain rule
\begin{equation}
\frac{\partial\bnabla_\w}{\partial\lambda}=
\HH_2^T
\frac{\partial\w}{\partial\lambda}+
\HH_1^H
\frac{\partial\w^*}{\partial\lambda}\label{ddwdl},
\end{equation}
where $\HH_2\eqdef\frac{\partial\bnabla_\w^T}{\partial\w}=\frac{\partial^2\mathcal{C}}{\partial\w^H\partial\w}$ and 
$\HH_1\eqdef\frac{\partial\bnabla_\w^T}{\partial\w^*}=\frac{\partial^2\mathcal{C}}{\partial\w^T\partial\w}$ are 
Hessian matrices corresponding to the conventional (unstructured) ICE model. 

Instead of the exact value of the second derivative $\frac{\partial^2\mathcal{C}}{\partial \lambda^2}$, we will now consider its analytic value when $\lambda$ and $\w$ are equal to their respective true values and $N=+\infty$. For that case, $\bnabla_\w=0$, hence only the second term in \eqref{secondderivative} is nonzero. We can use the results of Proposition~1 in \cite{koldovsky2021fastdiva} saying that, for that case, the values of the Hessian matrices are given by 
\begin{align}
    {\bf H}_1&=(c_3{\bf a}{\bf a}^T)^*,\label{H1}\\
    {\bf H}_2&=(c_1{\bf C}_\x+c_2{\bf a}{\bf a}^H)^T\label{H2},
\end{align}
where $c_1 = \frac{1}{\sigma^2}\left(\frac{\nu-\rho}{\nu}\right)$, $c_2 = -{\sigma^2}c_1-c_3$,
$c_3 = \frac{1}{2\nu}(\xi-\eta-\nu)$, $\rho = {\rm E}\left[\frac{\partial\phi(\frac{s}{\sigma})}{\partial s^*} \right]$, $\xi = {\rm E}\left[\frac{\partial\phi(\frac{s}{\sigma})}{\partial s^*}\frac{|s|^2}{\sigma^2} \right]$, and $\eta = {\rm E}\left[\frac{\partial\phi(\frac{s}{\sigma})}{\partial s}\frac{s^2}{\sigma^2} \right]$. What is left to compute is 
\begin{equation}\label{dwdl}
 \frac{\partial\w}{\partial \lambda}=\mi\sigma^2{\bf C}_\x^{-1}(\a\odot{\bf v})+2\w\Im\{\w^H(\a\odot{\bf v})\}.
\end{equation}
By putting \eqref{H1}-\eqref{dwdl} into \eqref{ddwdl} and then into \eqref{secondderivative} and assuming that $c_3=0$ (let us assume that the chosen nonlinearity $\phi(\cdot)$ satisfies this), the second derivative obtains the form
\begin{equation}\label{secondderivativeapprox}
\frac{\partial^2\mathcal{C}}{\partial \lambda^2}=
2c_1\sigma^2\left(\sigma^2 (\a\odot{\bf v})^H{\bf C}_\x^{-1}(\a\odot{\bf v}) - |\w^H(\a\odot{\bf v})|^2\right).
\end{equation}

We now propose the algorithm for finding the estimate of $\lambda$ as follows: Starting from an initial guess $\lambda=\lambda_{\rm ini}$, the algorithm repeats the following main steps
\begin{enumerate}
    \item compute $\a(\lambda)$ using \eqref{structureda},
    \item compute $\w(\lambda)$ using \eqref{orthogonalconstraint} and $s=\w^H\x$,
    \item evaluate the sample-based estimates of $\nu$, $\rho$, $\sigma^2$,
    \item perform update 
    $
        \lambda\leftarrow\lambda-\frac{\partial\mathcal{C}}{\partial \lambda}/\frac{\partial^2\mathcal{C}}{\partial\lambda^2},
    $
    where $\frac{\partial\mathcal{C}}{\partial \lambda}$ is given by \eqref{firstderivative} and $\frac{\partial^2\mathcal{C}}{\partial\lambda^2}$ is given by \eqref{secondderivativeapprox} where the quantities are replaced by their current sample-based estimates,
\end{enumerate}
until convergence (the change of the value of $\w$ is smaller than a threshold). The algorithm will be denoted as CaponICE.

\section{Cram\'er-Rao-induced bound on Interference-to-Signal Ratio}
The Interference-to-Signal Ratio (ISR) measured in the extracted signal $s=\w^H\x$ is equal to the ratio of variances of the SOI and the residual background in it. Its mean value is a relevant quality criterion for evaluating algorithms. The mean ISR is equivariant under the determined mixing model, which means that its value is independent of the true value of the parameter vectors $\a$ and $\w$ \cite{kautsky2020CRLB}. By Eq.~24 in \cite{kautsky2020CRLB}, 
\begin{equation}\label{CRIBgeneral}
    {\rm E}[{\rm ISR}]\geq\frac{1}{\sigma^2}{\tt tr}\{{\bf C}_\z{\tt CRLB}({\bf h})|_{{\bf h}=0}\},
\end{equation}
where ${\bf h}$ is the lower part of $\w$, i.e., $\w=[\beta,\,{\bf h}^T]^T$, ${\tt CRLB}\{\cdot\}$ denotes the Cram\'er-Rao Lower bound on the estimation variance of the argument, and ${\tt tr}\{\cdot\}$ denotes the trace. The right-hand side of \eqref{CRIBgeneral} is an algorithm-independent lower bound induced by the CRLB derived based on the equivariance of ISR. The value of this bound depends only on the model and reveals its fundamental limitations. For the conventional (unstructured) ICE, this bound says that \cite{hyvarinen1997b,loeschCRB,kautsky2020CRLB}
\begin{equation}\label{CRIBICE}
    {\rm E}[{\rm ISR}]\geq\frac{1}{N}\frac{d-1}{\overline\kappa-1},
\end{equation}
where $\kappa={\rm E}[|\psi(s)|^2]$ where $\psi(s)$ is the score function corresponding to the true pdf of the SOI, and $\overline\kappa=\kappa\sigma^2$. It holds that $\overline\kappa\geq 1$ and $\overline\kappa=1$ if and only if the SOI is circular Gaussian. Hence, the denominator in \eqref{CRIBICE} clearly shows that the model cannot identify circular Gaussian SOI. 

We now derive the bound for the proposed structured model using the results in \cite{kautsky2020CRLB}. By Eq.~28 in \cite{kautsky2020CRLB} and assuming the Gaussianity of the background signals, the log-likelihood function for one signal sample is given by 
\begin{equation}
    \mathcal{L}({\bf h},\lambda|\x)=\log p(\w^H\x)-\x^H{\bf B}^H{\bf C}_\z{\bf B}\x +\text{const.},
\end{equation}
where $p(\cdot)$ denotes the true pdf of the SOI. Owing to the equivariance of ISR, we can consider values of the derivatives of $\mathcal{L}$ only for ${\bf h}=0$ and $\lambda=0$, which gives
\begin{align}
    \frac{\partial\mathcal{L}}{\partial{\bf h}}&=\psi(s)\z, \\
    \frac{\partial\mathcal{L}}{\partial{\lambda}}&= 2\Im\{s\z^H{\bf C}_\z^{-1}{\bf v}\}.
\end{align}
Since ${\bf h}$ is complex-valued while $\lambda$ is real-valued, we compute the Fisher information matrix (FIM) for the mixed case parameter vector $\boldsymbol{\theta}=[{\bf h}^T,\,{\bf h}^H\,,\lambda]^T$ given by \cite{menni2012}
\begin{equation}\label{FIM}
    {\bf F}={\rm E}\left[\frac{\partial\mathcal{L}}{\partial \boldsymbol{\theta}}\left(\frac{\partial\mathcal{L}}{\partial \boldsymbol{\theta}}\right)^H\right]=\begin{pmatrix}
        \kappa{\bf C}_\z & 0 & -\mi\tilde{\bf v} \\
        0 & \kappa{\bf C}_\z^* & \mi\tilde{\bf v} \\
        \mi\tilde{\bf v}^T & -\mi\tilde{\bf v}^T & 2\sigma^2\tilde{\bf v}^T{\bf C}_z^{-1}\tilde{\bf v}
    \end{pmatrix},
\end{equation}
where we exploit the circularity of $\z$, i.e., ${\rm E}[\z\z^T]=0$; $\tilde{\bf v}=[v_2,\dots,v_d]^T$. Now, ${\tt CRLB}({\bf h})$ is obtained as the upper left-corner block of $N^{-1}{\bf F}^{-1}$, whose analytic value can be computed using block matrix inversion identity \cite{petersen2008} (we skip details due to lack of space). By putting the result into \eqref{CRIBgeneral}, it can be shown that
\begin{equation}\label{CRIBnewmodel}
    {\rm E}[{\rm ISR}]\geq\frac{1}{N}\left(\frac{d-1}{\overline\kappa}+\frac{1}{2}\frac{1}{\overline\kappa(\overline\kappa-1)}\right).
\end{equation}

The bound confirms the equivariance of ISR as it is independent of ${\bf C}_\z$ and $\sigma^2$, and it does not even depend on $\tilde{\bf v}$. Since $\overline\kappa\geq 1$ and $d>1$, it holds that the bound of the proposed mixing model given by \eqref{CRIBnewmodel} is lower than that of the unstructured model \eqref{CRIBICE}. As a result, higher extraction accuracy can be achieved with the proposed model if the data obeys it.


\section{Experimental Validation}
\subsection{  Simulation}
We now verify the performance of CaponICE and compare it with the lower bound given by \eqref{CRIBnewmodel} denoted as CaponCRiB. We also compare the general-mixture ICE model represented by the one-unit FastICA\footnote{We use the implementation from \cite{koldovsky2022double} with $K=T=L=1$.} algorithm \cite{hyvarinen1997b} and the corresponding bound \eqref{CRIBICE} denoted as ICECRiB. CaponICE and FastICA are implemented with the rational nonlinearity $\phi(s)=\frac{s^*}{1+|s|^2}$ \cite{rati2007,koldovsky2019TSP}. To compare conventional array processing approaches, we consider DOA estimation methods Root MUSIC and TLS ESPRIT to estimate $\lambda$ and evaluate the performance obtained through \mbox{${\bf w}(\lambda)$}.

The solution of BSE is known to be ambiguous in that any independent source can play the role of the SOI. The evaluation should always take into account whether the algorithm has converged to the desired source. Therefore, besides the output signal-to-interference ratio (SIR), we evaluate the success rate, which is equal to the percentage of trials where the resulting SIR is higher than $3$dB. The SIR is then averaged over the ``successful'' trials; all experiments are repeated in $1000$ trials. The blind algorithms are initialized randomly in the vicinity of the SOI; the initial performance is denoted as ``ini''. 

In a trial, a complex-valued mixture of $d$ Laplacean independent signals of length $N$ is generated as $\x(n)=\A{\bf u}(n)$ where $\A$ is $d\times d$, and its elements are random complex numbers of magnitude one. The powers of ${\bf u}(n)$ are set so that the SIR on input channels (iSIR) varies between $-20$ through $20$~dB. 
{  The first and second columns of $\A$ are $\a(\lambda_*)$ and $\a(\tfrac{1}{4})$, respectively, where $\a(\cdot)$ is defined in \eqref{structureda}.} 
Therefore, the first signal in ${\bf u}(n)$ plays the role of the SOI, and the second signal is a competitive interferer whose mixing vector also obeys \eqref{structureda}; the other $d-2$ interferers do not obey \eqref{structureda} in general so they are less competitive within the phase-shift mixing model (but fully competitive within the conventional ICE model). We consider a uniform linear array (ULA) of sensors where ${\bf v}=[0,1,2,\dots,d-1]^T$.

\begin{figure}
     \centering
     \includegraphics[width=0.9\linewidth]{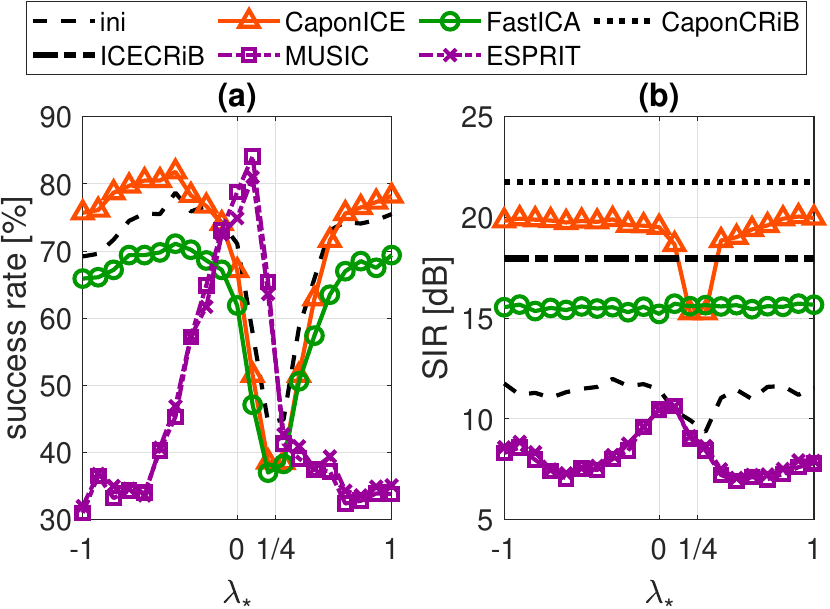}\\
     \caption{Global convergence and average SIR as functions of the ground true value $\lambda_\star$ corresponding to the SOI.}
     \label{fig:lambda}
\end{figure}

Figures~\ref{fig:lambda}(a) and (b) show the success rate and average SIR for $d=5$ and $N=500$ when $\lambda_*$ varies from $-1$ through $1$. The success rate shows that the region of convergence of CaponICE is much broader than that of FastICA. This is because FastICA can be attracted by all $d$ sources in the mixture. For $\lambda_*\approx\frac{1}{4}$, the success rate of both algorithms declines because the algorithms can converge to the second source with a higher probability. The SIR averaged over the successful trials does not depend on $\lambda_*$ up to small deviations for the critical $\lambda_*\approx\frac{1}{4}$. This confirms the equivariance of the algorithms (the performance neither depends on the mixing matrix nor on the iSIR). 

CaponCRiB is smaller than ICECRiB (in terms of SIR), confirming the higher achievable accuracy by the proposed model. CaponICE is comparably more accurate than FastICA, and both algorithms obey their respective CRiBs. The algorithms cannot attain the CRiBs because $\phi$ does not correspond to the true score function of the SOI.

The conventional DOA+beamforming methods MUSIC and ESPRIT show different behavior in terms of success rates than the blind methods and significantly lower SIR. Their success rates are lower than 40\% for $|\lambda|>1/4$ but increase significantly around $\lambda=0$. This phenomenon explains the fact that the methods are not able to identify the SOI in many trials (especially when iSIR$<0$dB) and tend to prefer the look-ahead direction ($\lambda=0$) in those cases.

    




\subsection{Frequency-domain speaker extraction}
We consider the cocktail party problem in a simulated $[6\,6\,2]$~m room (width-length-height) with the reverberation time set to T$_{60}=100$, $150$, and $200$~ms, respectively {  \cite{allen1979,sena2015}}. Four speakers at positions $[1.5\,3\,1]$, $[2\,3\,1]$, $[3\,3\,1]$, and $[4\,2\,1]$ utter simultaneously for $5$ seconds and are recorded by $5$ omnidirectional microphones arranged in a uniform linear array (parallel to x-axis) with $5$~cm spacing; the central microphone is placed at $[2\,1\,1]$. The signals are sampled at $16$~kHz and transformed by the short-term Fourier transform with the FFT length $1024$ and shift of $128$ samples.

We consider a straightforward modification of the proposed method for joint processing of all frequency bands based on IVE \cite{kim2006,koldovsky2019TSP}. Here, $\lambda$ is the joint mixing parameter for all frequencies, and the contrast function is equal to the averaged value of \eqref{contastICE} over the frequencies. A joint non-linear function for the $k$th frequency is selected as
$\phi_k(s_1,\dots,s_k)=\frac{s_k^*}{1+\sum_k|s_k|^2}$. The first and second derivatives of the contrast function are, respectively, equal to the averages of \eqref{firstderivative} and \eqref{secondderivativeapprox} over the frequencies. To compare, the speakers' DOAs are estimated by finding the corresponding local maxima of the steered-response power with phase transform (SRP-PHAT) using {\tt fminsearch} in Matlab. The DOA of each speaker is estimated by initializing the methods by the true angle$+5^{\circ}$. 

Table~\ref{tab:angles} shows that the estimated DOA deteriorates with increasing T$_{60}$. The emphasized values indicate cases where the method is no longer able to identify a given speaker and estimates the DOA of another speaker. The proposed method shows very accurate DOA estimation in the lowest reverberation environment and more robustness at longer reverberation times. The improvement of SIR of the extracted speakers by the proposed method {  (SRP-PHAT) is 7.0 (6.7), 2.2 (2.0), 8.0 (8.0), 11.2 (11.0) dB (100 ms); 5.3 (1.1), 2.2 (2.1), 6.9 (-1.9), 7.2 (5.5) dB (150 ms); and 4.5 (1.9), -1.0 (1.6), 5.1 (-2.0), 6.0 (4.5) dB (200 ms)}.

\begin{table}
    \caption{Estimated DOAs of speakers}
    \centering
    \begin{tabular}{l|c|c|c|c}
        method\textbackslash speaker & 1 & 2 & 3 & 4 \\\hline
        true DOA ($^{\circ}$)& 104.04 & 90.00 & 63.43 & 26.57 \\
        SRP-PHAT $100$ ms & 103.01 & 91.28 & 63.89 & 23.63 \\
        SRP-PHAT $150$ ms & {\it 91.27} & 91.27 & {\it 91.27} & 22.41 \\
        SRP-PHAT $200$ ms & {\it 91.20} & 91.20 & {\it 91.20} & 22.93 \\
        proposed $100$ ms & 104.30 & 90.02 & 63.77 & 26.85 \\
        proposed $150$ ms & 102.69 & 90.08 & 67.87 & 30.75 \\
        proposed $200$ ms & 102.12 & {\it 102.07} & 69.74 & 34.00
    \end{tabular}
    \label{tab:angles}
\end{table}

\section{Conclusions}
In this letter, we have shown that ICE and IVE can be implemented for a structured mixing model with a significantly reduced number of mixing parameters. 
The derived method is, naturally, limited by the validity of the model. The simulations and the case study of speaker extraction have shown that the model yields benefits as long as the observed data behaves according to the model or does not deviate too much. In future research, we plan to focus on semi-blind models decomposable into trainable structures \cite{shlezinger2023}, which have the potential to be more robust to mixing model violations and to take advantage of information from training examples.



\balance

\end{document}